\documentclass[aps,prl,twocolumn,showpacs,superscriptaddress,groupedaddress]{revtex4} 
\usepackage{graphicx}  
\usepackage{dcolumn}   
\usepackage{bm}        
\usepackage{amssymb}   
\usepackage{amsmath}
\usepackage[english]{babel}
\usepackage[utf8]{inputenc}
\begin{document}
\title{Molecular beam depletion: a new approach}
\begin{abstract}
During the last years some interesting experimental results have been reported for experiments in N$_2$O, NO, NO dimer, H$_2$, Toluene and BaFCH$_3$ cluster. The main result consists in the observation of molecular beam depletion when the molecules of a pulsed beam interact with a static electric or magnetic field and an oscillating field (RF). In these cases, and as a main difference, instead of using four fields as in the original technique developed by I.I. Rabi and others, only two fields, those which configure the resonant unit, are used. That is, without using the non-homogeneous magnetic fields. 

The depletion explanation for I.I. Rabi and others is based in the interaction between the molecular electric or magnetic dipole moment and the non-homogeneous fields. But, obviously, the change in the molecules' trajectories observed on these new experiments has to be explained without considering the force provided by the field gradient because it happens without using non-homogeneous fields.

In this paper a theoretical way for the explanation of these new experimental results is presented. One important point emerges as a result of this development, namely, the existence of an until now unknown, spin-dependent force, which would be responsible of the aforementioned deviation of the molecules.

\end{abstract}


\author{Manuel Dorado}
\email{mdorado@cirta.es}
\affiliation{CIRTA. Rotation and Torque Research Center. Theoretical group.\\
C$/$Maratón, 6. C.P. 28037, Madrid, Spain.}

\maketitle
\section{Introduction}
\begin{figure}[h]
\includegraphics[scale=0.43]{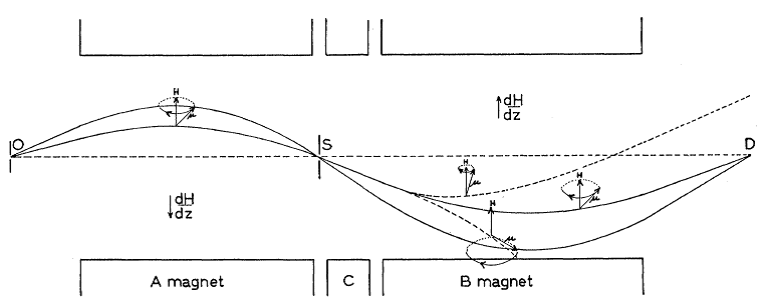}
\caption{\label{fig:1} Typical path of molecules in a M.B.M.R. experiment. The two solid curves show the paths of the molecules whose moments do not change when passing through the resonant cell.}
\end{figure}

The molecular beam magnetic resonance (MBMR) technique has significantly contributed, as is well known, to the development of atomic and molecular physics \cite{ramsey}. And it makes possible to measure the Larmor frequency of an atom or molecule in the presence of a magnetic field. In the original technique, developed by I.I. Rabi and others \cite{rabi1}, \cite{rabi2} the molecular beam is forced to pass through four different fields:
A non-homogeneous polarizer field (A) where the molecules are prepared.
A resonant unit (C) that consists of two, a static and an oscillating, fields. 
A non-homogeneous analyzer field (B). Only molecules in the prepared state reach the detector. 
The two non-homogeneous magnetic fields A and B have opposite directions.
The molecular beam describes a sigmoidal trajectory and, finally, is collected in a detector (see fig. \ref{fig:1}).

Rabi explained this effect in terms of spatial reorientation of the angular moment due to a change of state when the transition occurs.
In this case the depletion explanation is based on the interaction between the molecular magnetic dipole moment and the non-homogeneous fields. 
\begin{equation}
\vec{F} = \nabla \left(\vec{\mu}_z \vec{B}\right)                                                                    
\label{eq:1}
\end{equation}                                                                                                 

The force is provided by the field gradient interacting with the molecular dipolar moment (electric or magnetic).
On the resonant unit the molecular dipole interacts with both, homogeneous and oscillating, fields. When the oscillating field is tuned to a transition resonant frequency between two sub states, a fraction of the molecular beam molecules is removed from the initial prepared state. The dipolar moment changes in this fraction and as a consequence, the interaction force with the non-homogeneous analyzer field (B). As only molecules in the initial prepared state reach the detector the signal in the detector diminishes.

\section{New experimental results}

\begin{figure}[h]
\includegraphics[scale=0.55]{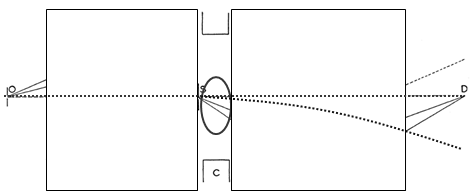}
\caption{\label{fig:2} The dotted line path show the trajectory change of the fraction of the molecular beam that is removed from the initial prepared state when passing through the resonant cell.}
\end{figure}

During the last years some interesting experimental results have been reported for N$_2$O, NO, NO dimer, H$_2$ and BaFCH$_3$ cluster \cite{urena} - \cite{montero2}. The main result consists in the observation of molecular beam depletion when the molecules of a pulsed beam interact with a static electric or magnetic field and an oscillating field (RF) as in the Rabi's experiments.
But, in these cases, instead of using four fields, only two fields, those which configure the resonant unit (C), are used, that is, without using the non-homogeneous magnetic, A and B, fields. See fig.\ref{fig:2}
 
In a similar way, when the oscillating field is tuned to a transition resonant frequency between two sub states, the fraction of the molecular beam that is removed from the initial prepared state does not reach the detector. But the important thing is: differently to the previous method, it happens without using non-homogeneous fields. 
Obviously, the trajectory change has to be explained without considering the force provided by the field gradient.
There must be another molecular feature that explains the depletion. It looks as though the linear momentum conservation principle were not satisfied. These experiments suggest that a force depending on other fundamental magnitude of the particle, different from mass and charge must be taken into account.

\subsection{Looking for an explanation...}

In order to find out an explanation, let's consider the following case:
An electron is moving, with speed, $\vec{v}$ constant in modulus, in a homogeneous magnetic field  $\vec{B}$  where $\vec{v}$ is perpendicular to $\vec{B}$.
Its kinetic energy will be:

\begin{equation}
E = \frac{1}{2}mv^2                                                                  
\label{eq:2}
\end{equation}
                                                                                
The electron, as is well known, describes a circular trajectory (in general case an helix) with a radius $r$, being:
\begin{equation}
r = \frac{v}{\omega}                                                                  
\nonumber
\end{equation}
and:
\begin{equation}
\vec{\omega}= \frac{q\vec{B}}{m}                                                                  
\label{eq:3}
\end{equation}

due to the Lorentz force:
\begin{equation}
\vec{F} = q\vec{v}\times \vec{B}                                                                  
\label{eq:4}
\end{equation}
                                                                                                        
On the other hand, as the electron has a magnetic moment, $\vec{\mu}$, and spin $\vec{S}$, the presence of the magnetic field $\vec{B}$ produces a torque when interacting with the electron magnetic moment $\vec{\mu}$. The angle between $\vec{S}$ and O$_z$ (the direction of the magnetic field  $\vec{B}$) remains constant but the spin $\vec{S}$ revolves about  O$_z$ with angular velocity $\vec{\Omega}$. This phenomenon bears the name of Larmor precession.
The electron kinetic energy must increase with the energy due to spin precession.
But it should be considered that the forces producing the torque are perpendicular to the precession motion and, as a consequence, do not modify the energy of the system.  
It looks like if the principle of energy conservation be violated. 

\section{How to solve this dilemma?}

\subsection{First option}
If the rotation around an axis is considered as origin of the spin, in a classic (and impossible) interpretation, one could imagine the electron rotating in a slowly way and offsetting the increase in energy due to the precession movement.
But, as it is well known, the spin is a quantized quantity; its modulus is constant and immutable. 
This option is, as a consequence, not acceptable.

\subsection{Second option.}
Let us consider now that the helicity is a constant of motion. 
Helicity, $\xi$, is defined as the scalar product of linear momentum and the spin: 
\begin{equation}
 \xi = \left(m\vec{v}\right) \cdot \vec{S}                                                               
\label{eq:5}
\end{equation}
                                                                  
Is this hypothesis consistent with Quantum Mechanics?     
Let us consider an electron in a uniform magnetic field $\vec{B}$, and let us choose the O$_z$ axis along $\vec{B}$.
The classical potential energy due to electron magnetic moment $\vec{\mu}$ is then
\begin{equation}
\label{eq:6}
 W = - \vec{\mu} \cdot \vec{B} = -\mu_{z} B = -\gamma B S_{z} 
\end{equation} 
                                                                          
where $B$ is the modulus of the magnetic field. Let us set:
\begin{equation}
\omega_0 = -\gamma B = \Omega                                                                  
\label{eq:7}
\end{equation}
                                                                                                    
$\Omega$ being the classical angular precession velocity. 
(As is well known, $\omega_0$ has dimensions of the inverse of a time, that is, of an angular velocity.)
If we replace $S_z$  by the operator $S_z$ the classic energy becomes an operator: the Hamiltonian $H$ which describes the evolution of the spin of the electron in the field $\vec{B}$ is:

\begin{equation}
H = \omega_0 S_z                                                                 
\label{eq:eq8}
\end{equation}

Since this operator is time independent, solving the corresponding Schr$\ddot{o}$dinger equation amounts to solving the eigenvalue equation of H. We immediately see that the eigenvectors of H are those of $S_z$ (see ref.\cite{cohen}):

\begin{equation}
H \left|+\right\rangle = + \frac{\hbar \omega_0}{2} \left|+\right\rangle 
\label{eq:9}
\end{equation}

\begin{equation}
H \left|-\right\rangle = - \frac{\hbar \omega_0}{2} \left|-\right\rangle 
\label{eq:10}
\end{equation}
                                                            
There are therefore two energy levels, $E_+= + \frac{\hbar \omega_0}{2}$ and $E_-= - \frac{\hbar \omega_0}{2}$
Their separation $\hbar \omega_0$ is proportional to the magnetic field and define a \textit{single Bohr frequency}
\begin{equation}
\nu_{+-} = \frac{1}{h}\left(E_+ - E_-\right) = \frac{\omega_0}{2\pi}                                                              
\label{eq:eq11}
\end{equation}
                                                                                     
Is it possible to distinguish, in a uniform magnetic field $\vec{B}$, which electrons are the state $\left|+\right\rangle$ and which are the state $\left|-\right\rangle$? The answer is no. Their behavior inside the field is exactly the same.
But, nevertheless, if we introduce a oscillating magnetic field H$_1$ with a frequency resonant with the transition 
$\nu_{(+-)}=\frac{1}{\hbar}\left(E_+ + E_-\right) = \frac{\omega_0}{2\pi}$, then  it will be possible to distinguish both states by the difference in their trajectories, see ref.\cite{dorado2}. 

\section{Larmor Precession}

Let us assume that, at time $t = 0$, the spin is in the state:
\begin{equation}
\left|\chi (0) \right\rangle = \text{cos} \frac{\theta}{2} e^{\frac{-i\phi}{2}}\left|+\right\rangle + \text{sin} \frac{\theta}{2} e^{\frac{i\phi}{2}}\left|-\right\rangle                                                                 
\label{eq12}
\end{equation}
                                                    
To calculate the state $\left|\chi (t) \right\rangle$ in an arbitrary state $t > 0$ and as $\left|\chi (0) \right\rangle$ is already expanded in terms of the eigenstates of the Hamiltonian we will obtain: 

\begin{equation}
\left|\chi (t) \right\rangle = \text{cos} \frac{\theta}{2} e^{\frac{-i\phi}{2}}e^{-iE_+ \frac{t}{\hbar}}\left|+\right\rangle + \text{sin} \frac{\theta}{2} e^{\frac{i\phi}{2}}e^{-iE_- \frac{t}{\hbar}}\left|-\right\rangle                                                                
\label{eq:13}
\end{equation}

Or, using the values of E$_+$ and E$_-$:
\begin{equation}
\left|\chi (t) \right\rangle = \text{cos} \frac{\theta}{2} e^{+\frac{-i\left(\phi + \omega_0 t\right)}{2}}\left|+\right\rangle + \text{sin} \frac{\theta}{2} e^{-\frac{i\left(\phi + \omega_0 t\right)}{2}}\left|-\right\rangle                                                                
\label{eq:14}
\end{equation}

The presence of the magnetic field $\vec{B}$ therefore introduces a phase shift, proportional to time, between the coefficients of the kets $\left|+\right\rangle$ and $\left|-\right\rangle$.

Comparing the expression (\ref{eq:14}) for $\left|\chi (t) \right\rangle$ with that for the eigenket $\left|+\right\rangle_\mu$  for the observable $\vec{s}\cdot \vec{u}$ 

\begin{equation}
\left|+ \right\rangle _{u} = \text{cos} \frac{\theta}{2} e^{\frac{-i\phi}{2}}\left|+\right\rangle + \text{sin} \frac{\theta}{2} e^{\frac{i\phi}{2}}\left|-\right\rangle   
\label{eq:15}
\end{equation}

We see that the direction $u(t)$ along which the component is $+\frac{\hbar}{2}$,with certainty, is defined by the polar angles:

\begin{eqnarray}
\nonumber
\theta_t &=& \theta = \text{Cte} \\
\phi(t) &=& \phi + \omega_0 t                                                             
\label{eq:16}
\end{eqnarray} 
                                                                                              
The angle between $u(t)$ and O$_z$ (the direction of the magnetic field $\vec{B}$) therefore remains constant, but  $u(t)$ revolves around O$_z$ with angular velocity  $\vec{\omega_0} = \vec{\Omega}$ proportional to the magnetic field.
Thus, we find in quantum mechanics the phenomenon equivalent to that described for a particle with classic magnetic moment and spin and which bears the name of Larmor precession.

\section{Helicity as a constant of motion}

We redefine now the helicity, $\xi$, in order that its eigenvalues be $\pm$1, as $\xi = \vec{\sigma}\cdot \hat{v}$, where $\hat{v} = \frac{\vec{v}}{v}$ and $\vec{s}=\frac{\hbar}{2}\vec{\sigma}$. The initial velocity of the electron is $\vec{v} = v_0 \left(\cos \phi, \sin \phi, 0\right)$, and we assume the initial spin state of the electron to be an eigenstate of the helicity with eigenvalue +1, which is given in (\ref{eq12}), with $\theta = \frac{\pi}{2}$, that is:
\begin{equation}
\left|\chi (0) \right\rangle = \frac{\sqrt{2}}{2} \left(e^{\frac{-i\phi}{2}}\left|+\right\rangle + e^{\frac{i\phi}{2}}\left|-\right\rangle\right)                                                             
\label{eq:17}
\end{equation}
At the time $t$ the velocity of the electron is, as it is known,
\begin{equation}
v\left(t\right) = v_0 \left[\cos\left(\phi + \omega_o t\right), \sin\left(\phi + \omega_0 t\right),0\right]                                                           
\label{eq:18}
\end{equation}
where $\omega_0$ is given in (\ref{eq:7}).
According to (\ref{eq:15}) and (\ref{eq:16}), with $\theta = \frac{\pi}{2}$, at time $t$ the spin state is:
\begin{equation}
\left|\chi (t) \right\rangle = \frac{\sqrt{2}}{2} \left(e^{-\frac{i\left(\phi + \omega_0 t\right)}{2}}\left|+\right\rangle + e^{\frac{i\left(\phi + \omega_0 t\right)}{2}}\left|-\right\rangle\right)                                                             
\label{eq:19}
\end{equation}
and the helicity at time $t$, $\xi(t) = \vec{\sigma}\cdot \hat{v}(t)$,

\begin{equation}
\xi\left(t\right) = \left[\sigma_x \text{cos}\left(\phi + \omega_0 t \right), \sigma_y\text{sin}\left(\phi + \omega_0 t\right),0\right]                                                           
\label{eq:20}
\end{equation}
Now, taking into account that,
\begin{equation*}
\sigma_x \left(\left|+\right\rangle,\left|-\right\rangle\right) =\left(\left|-\right\rangle,\left|+\right\rangle\right);\sigma_y \left(\left|+\right\rangle,\left|-\right\rangle\right) =\left(i\left|-\right\rangle,-i\left|+\right\rangle\right) 
\end{equation*}
we easily obtain:
\begin{equation}
\xi(t) \left|\chi(t)\right\rangle  = \left|\chi(t)\right\rangle                                                          
\label{eq:21}
\end{equation}
This shows that $\left|\chi(t)\right\rangle$ is an eigenstate of the helicity of eigenvalue +1; in other words, helicity is conserved along the electron's (classical) trajectory.

\section{Consequences}

It has been proven that helicity is a constant. As a consequence of this result, the linear momentum $m\vec{v}$ must have the same precession angular velocity (Larmor angular velocity) $\vec{\Omega}$ than the spin $\vec{S}$. The equation of motion describing the linear momentum evolution must be then equivalent of the equation of motion which describe the evolution of the spin $\vec{S}$. This means that:

\begin{equation}
\left(\frac{d\vec{S}}{dt}\right)_{\text{Inertial}} = \vec{S} \times \vec{\Omega}
\label{eq:22}
\end{equation}

\begin{equation}
\left(\frac{dm\vec{v}}{dt}\right)_{\text{Inertial}} = m \vec{v} \times \vec{\Omega}
\label{eq:23}
\end{equation}
                                                                                            
It is concluded the particle will be under a central acceleration, $\vec{a} = \vec{v} \times \vec{\Omega}$  perpendicular to $\vec{v}$. The particle is then under a central force: 
\begin{equation}
\vec{F} = m \vec{v} \times \vec{\Omega}
\label{eq:24}
\end{equation}
This kind of forces related with the spin will be designed as Lorentz-like forces.                                                                                                             	
In this case, the trajectory will be a circular one. The radius will be: 
\begin{equation}
R = \frac{\left|\vec{v}\right|}{\left|\vec{\Omega}\right|}
\label{eq:25}
\end{equation}
And its kinetic energy:  
\begin{equation}
E = \frac{1}{2}I_z \Omega^2 = \frac{1}{2}m R^2 \Omega^2 = \frac{1}{2}mv^2
\label{eq:26}
\end{equation}
which is equal to the initial one shown in (\ref{eq:2}).
The force (\ref{eq:24}) is the responsible of the electron circular trajectory inside the field  $\vec{B}$ and should be related to the spin $\vec{S}$  of the electron.

\subsection{Electron in a magnetic field}

If the case of an electron in a magnetic field is considered, then the force due to the spin of the electron will be:

\begin{equation*}
\vec{F} = m \vec{v} \times \vec{\Omega}
\end{equation*}

Where $\vec{\Omega}$ is the spin Larmor precession velocity around O$_z$.
But is known that:

\begin{equation}
\vec{\Omega} = \frac{\left|\vec{\mu}\right|\vec{B}}{\left|\vec{S}\right|} = \frac{gq\vec{B}}{2m}
\label{eq:27}
\end{equation}
                                  
Substituting in (\ref{eq:24}) the expression for the force acting on the particle is obtained. This force has its origin on the spin. This expression is:
\begin{equation}
\vec{F} = \frac{mg}{2m}q \vec{v} \times \vec{B}
\label{eq:28}
\end{equation}

As for an electron $g = 2$, the final result is: 
\begin{equation}
\vec{F} = q \vec{v} \times \vec{B}
\label{eq:29}
\end{equation}

Surprisingly this expression for the Lorentz-like force, related to the spin, coincide with that known as Lorentz force related to the charge. 
Considering the spin as responsible of the Lorentz-like force, a new deflection mechanism has been proposed (see ref. 10). The equations of motion for a system with intrinsic angular momentum when applying torques are described and, according with the theory, when the frequency of the oscillating field coincides with a transition resonant frequency (Larmor frequency), the molecules that change their state from the original one are removed from their trajectories and, as a consequence, do not reach the detector and the corresponding signal decreases. 

\section{New experimental proposal}

In 1939 Alvarez and Bloch \cite{bloch} measured the neutron magnetic moment by using a neutron beam passing through a resonant unit. Neutrons from the Be + D reactions were slowed to thermal velocities and diffused down a cadmium lined tube through the water tank to the polarizer magnet, B$_A$. After passing through the resonant unit that consists of two, a static and an oscillating, fields and the analyzer magnet,B$_B$ they were detected in a BF$_3$ chamber.
The polarizer B$_A$ and analyzer B$_B$ are strongly magnetized iron pieces. 
A neutron resonant dip is observed in the signal of the neutron beam when the oscillating resonant frequency corresponding to the transition between the two states up and down is reached. 
According to the previous theoretical description and recently results obtained for NO$_2$, NO, NO dimer, H$_2$ and BaFCH$_3$ cluster, if the Alvarez and Bloch experiment is carried out without using analyzer magnet B$_B$ , we anticipate that the experimental results will be the same as those obtained by Alvarez and Bloch in the experiment of 1939. 
According to the new explanation, the trajectory change takes place when neutrons pass through the resonant unit and the oscillating field is tuned to a transition resonant frequency between two states, up and down, of the spin of the neutron. 
In case of Alvarez and Bloch experiment, they used a magnetic field for the neutron resonance of 622 Gauss and a resonant frequency of oscillator of 1843 kilocycles.

\section*{Acknowledgments}
The author is very grateful to prof. José L. Sánchez Gómez, Universidad Autónoma de Madrid for useful discussions.

\end{document}